# IN VIVO MEASUREMENT OF BLOOD CLOT MECHANICS FROM COMPUTATIONAL FLUID DYNAMICS BASED ON INTRAVITAL MICROSCOPY IMAGES

Olufemi Emmanuel Kadri[1,*], Vishnu Deep Chandran[1,*], Surblyte, Migle[2,*], and Roman S. Voronov[1,†]

[1] Otto H. York Department of Chemical and Materials Engineering, New Jersey Institute of Technology, Newark, NJ 07102, USA

[2] Ying Wu College of Computing Sciences, Department of Computer Science, New Jersey Institute of Technology, Newark, NJ 07102, USA

[*] The first three authors contributed equally to the manuscript.

[†] Address correspondence to Roman S. Voronov, Otto H. York Department of Chemical and Materials Engineering, New Jersey Institute of Technology, Newark, NJ 07102, USA. Electronic mail: *rvoronov@njit.edu*, Fax: +1 973 596 8436, Tel: +1 973 642 4762

## Abstract

Ischemia leading to heart attacks and strokes is the major cause of deaths in the world. Whether an occlusion occurs or not, depends on the ability of a growing thrombus to resist forces exerted on its structure. This manuscript provides the first known *in vivo* measurement of the stresses that clots can withstand, before yielding to the surrounding blood flow. Namely, Lattice-Boltzmann Method flow simulations are performed based on 3D clot geometries. The latter are estimated from intravital microscopy images of laser-induced injuries in cremaster microvasculature of live mice. In addition to reporting the blood clot yield stresses, we also show that the thrombus "core" does not experience significant deformation, while its "shell" does. This indicates that the latter is more prone to embolization. Hence, drugs should be designed to target the shell selectively, while leaving the core intact (to minimize excessive bleeding). Finally, we laid down a foundation for a nondimensionalization procedure, which unraveled a relationship between clot mechanics and biology. Hence, the proposed framework could ultimately lead to a unified theory of thrombogenesis, capable of explaining all clotting events. Thus, the findings presented herein will be beneficial to the understanding and treatment of heart attacks, strokes and hemophilia.

# 1. Introduction

Achieving hemostasis following penetrating injuries is essential for the survival of organisms that possess a closed high-pressure circulatory system. However, pathological manifestation of thrombosis and embolism can potentially lead to life-threatening complications when occurring in the heart (i.e., a heart attack), brain (i.e., a stroke), or lungs (i.e., DVT/PE). Among these, thrombo-embolic infarction is the leading cause of mortality and morbidity in the United States, while stroke is the 5$^{th}$.[1]  Conversely, deficiencies in the clotting mechanisms (hemophilia or due to drug interactions) can result in bleeding risks that confront surgeons on a regular basis.  Yet, despite tremendous efforts by the medical research community (e.g., ~$3 billion of annual expenditure on heart attack and brain stroke research alone[2]), the problem that essentially amounts to a clogged "pipe" remains largely unsolved to this day.  Moreover, what makes one thrombus benign, while another one dangerous, is also not well understood.  Therefore, gaining insight into the thrombi's tendency to occlude blood vessels would be beneficial for public health, since it could pave the way towards a better understanding of the risk factors involved; and subsequently to better disease treatments and thromboectomy devices.[3]

## 1.1. Observed Structural Heterogeneity of Thrombi Imply Mechanical Differences

Whether an occlusion occurs or not, depends on the ability of the growing thrombus to resist the blood flow forces exerted on its structure. With development of advanced intravital microscopy experiments, the thrombi structures have been shown to be heterogeneous: consisting of a densely packed "core" nearest the injury site, and a loose "shell" overlaying the core (see **Figure 1**).[4, 5].  It is also reported that the core is composed of highly "activated" platelets (as is measured by P-selectin expression), while the shell consists of loosely-packed P-selectin -negative cells.  The biological purpose, as well as the cause of this heterogeneity, are unknown.  One thing that is apparent, however, is that the core and the shell contribute differently to key parts of the thrombus formation and hemostasis:  The shell is observed to shed the most mass (leading to the conclusion that embolism is mostly caused by this part of the clot); while the core can be seen to anchor the thrombus to the injury, and stop the escape of blood to the extravascular space by "sealing" the damage.  This leads to an important conclusion that there are potentially significant material and functional differences between these two regions of blood clots.

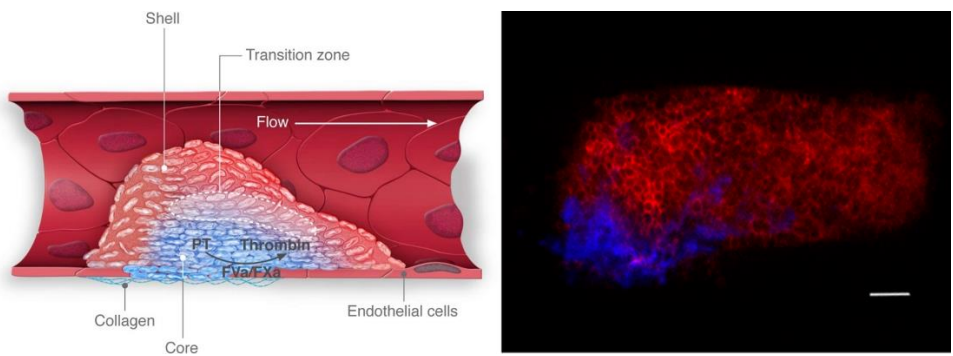

**Figure 1**  LEFT - "Core-and-Shell" model schematic showing that the clot is comprised of two regions differing in degree of platelet activation and packing density; Image reprinted with permission, from Ref.[6] RIGHT – Confocal fluorescent microscopy image of a clot (Blue = P-selectin exposure marking the activated core; Red = anti-CD41 platelet marker). Scale bar is 10µm.

## 1.2. Hypothesis: Bingham-like Viscoplastic Behavior of Blood Clots

At the same time, the viscoplastic behavior exhibited by the thrombi resembles that of a Bingham plastic - a material, like toothpaste, that behaves as a rigid body at low stresses, but flows as a viscous fluid when its critical yield stress $\sigma_c$ is exceeded (see Equations 1 – 2). Namely, like a Bingham plastic, the clot consists of discrete particles (in this case platelets) trapped in a liquid gel. The platelets interact with each other creating a weak solid structure known as a "false body". A certain amount of stress corresponding to $\sigma_c$ is required to break this structure and allow the platelets to rearrange within the gel under viscous forces. After the stress is reduced however, the platelets associate again, solidifying the structure. **Figure 2** illustrates the Bingham plastic-like behavior of a blood clot as observed from intravital microscopy.

$$\dot{\gamma} = 0 \qquad \sigma < \sigma_c, \text{ no flow} \qquad (1)$$

$$\sigma = \sigma_c + \mu\dot{\gamma}, \qquad \sigma \geq \sigma_c, \text{ flow with a constant viscosity} \qquad (2)$$

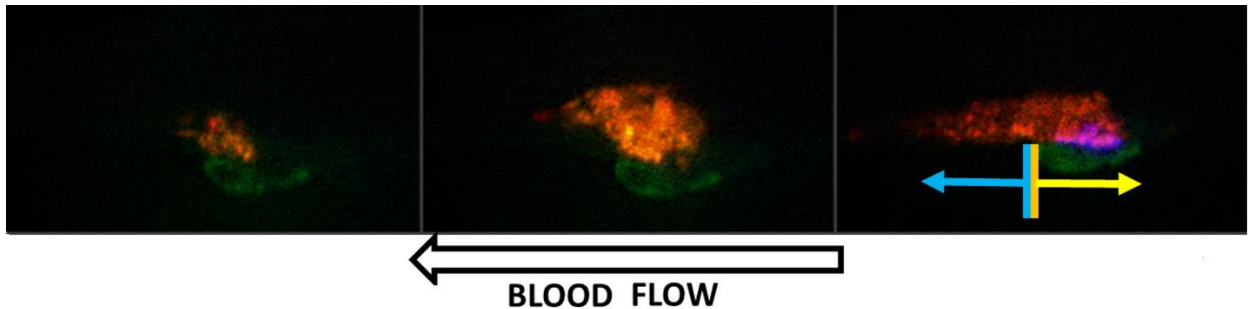

**Figure 2** Red – Platelets (Anti-CD41 platelet marker); Blue – Core (anti-P-selectin thrombus "core" marker); Green – Injury (Uncaged Fluorescent Albumin). White Arrow – indicates the direction of flow; Yellow Arrow – Upstream portion of the clot; Cyan Arrow – Downstream portion of the clot. A – Initial attachment at injury; B – Uniform growth prior to transition; C – Final stable steady-state core-and-shell thrombus structure, after the thrombus' yielding to the blood flow.

The figure shows three major progression stages of a typical thrombus formation: 1) initial platelet attachment to the injury site (see Figure 2-A); 2) clot growth radially outward from the injury site (see Figure 2-B); and 3) steady state stability (see Figure 2-C). The transition from the second to the third stage (Figure 2, panels B→C) is the Bingham-like "yielding" of the thrombus. During this critical event, platelet mass from the upstream portion of the thrombus is forced to its downstream side. As a result, the thrombus changes shape from a "ball"-like structure, to the characteristic "comet-tail" shape typically seen in blood clots. This transition occurs because the obstructing thrombus experiences stronger forces from the surrounding fluid, which is trying to squeeze through the little remaining openings left in the lumen. Consequently, if there is no full occlusion of the blood vessel, the thrombus' structure eventually *yields* to the flow forces, and rearranges its shape to minimize the fluid drag imposed on its surface.

Interestingly, despite the underlying complexity of the thrombo-genesis mechanism, the discrete regimes shown in Figure 2 appear to be common to all thrombi. Therefore, it is hypothesized here that the critical yield stress $\sigma_c$ is a parameter that is key to understanding the extent to which the clot's structure can resist deformation and breakage. In other words, it is an important measure of stiffness that can provide information on when the thrombus is likely to become pathogenic. Yet, $\sigma_c$ remains unmeasured to this day. This is because insight into thrombo-genesis is made difficult by the fact that it is a fast, small scale process that involves a combination of coupled biochemical reaction cascades, intra- and inter- cell signaling, cell and tissue biomechanics, and non-Newtonian fluid flow.[7, 8]

## 1.3. State-of-the-Art: Limitations of Stand-alone Experiments and Simulations

Among the experimental techniques, compression testing,[9, 10] tensile testing,[11-13] shear rheometry,[14, 15] nano-indentation[16] and ultrasound elastography[17-19] are commonly used to estimate mechanical properties of the thrombi (e.g., elastic modulus, shear modulus and stiffness).[9-12, 14, 16-18] However, most of these works use *in vitro* flow loops to generate thrombi, which may not be representative of the real physiological conditions *in vivo*. For example, there are some discrepancies between the *ex vivo* and the *in vivo* measurements. One such discrepancy is the platelet attachment, which happens so fast *in vivo*, that even the high speed cameras have trouble resolving it.[20] In contrast, the time scales of platelet activation observed *ex-vivo* are on the order of minutes, which is several orders of magnitude longer than that of *in-vivo*.[21-24]

Alternatively, a preformed thrombus could be explanted from the body for an *ex vivo* measurement. However, this type of experiment corresponds to just a single time point, and only at a late stage of the thrombus formation. Hence, it would not capture the full thrombo-genesis dynamics; knowledge of which is necessary to measure the critical yield stress of the clot. Unfortunately, only a few studies among the above works measure thrombi biomechanical properties directly *in vivo*. A typical example is Mfoumou et al.,[19] who used shear wave ultrasound imaging to measure thrombus stiffness in rabbits' veins. However, a) the temporal resolution of the measurements in such studies is on the order of $\Delta t = 10$ mins, which is again too slow to capture thrombus growth dynamics (lasts on the order of seconds), and b) the measurement occurs through tissue, which could reduce its accuracy. Therefore, better approaches are needed to deduce the ability of blood clots to resist deformation *in vivo*.

Computational models offer an attractive alternative, because they can resolve the time scale limitation by recreating thrombo-genesis *in silico*.[25-32] However, the clot structures generated by the simulations are not guaranteed to have realistic geometries, or to follow a realistic deformation trajectory over time. The reasons behind these limitations are the numerous unknowns in biology: for example, simulated clot structures are typically represented using either continuum models,[25, 27, 30, 32-35] discrete particle-based[26, 36, 37] or hybrid continuum/discrete particle-based[28, 29, 31, 38-40] dynamics. All these approaches rely on defining the clot biomechanics via parameters such as platelet-platelet, platelet-vessel wall/injury, platelet-proteins and fibrin-fibrin bond strengths (typically modeled using simple spring constants). However, since most of these parameters are obtained from *in vitro* experimental measurements, it is difficult to verify whether they are truly representative of the *in vivo* values. Moreover, some of the biological processes are simply too complex, and consequently require numerous simplifications / assumptions for achieving bottom-up modeling. For example, platelet

activation – a process central to clotting, is typically modeled using top-down neural networks pre-trained on an individual's unique platelet phenotypes;[41] while bottom-up approaches to modeling the same phenomenon require unknown-parameter estimation.[42] Therefore, the purely computational methods, ultimately do not guarantee a realistic trajectory of thrombus evolution over time.

## 1.4. Proposed Hybrid Image-based Modeling Approach, and 3D Clot Structure Estimation

Thus, this seemingly "simple" phenomenon that is responsible for a wide range of life-threatening pathologies is not easily accessible to either experimental or computational inquiries alone. Consequently, semi-empirical approaches offer a reasonable compromise for overcoming these limitations: they bypass the need for generating the clot structure mathematically, by obtaining it from experimental imaging instead. This ensures that a realistic geometry is used for solving the physics involved in the process. For example, in refs. [24, 30, 43, 44] such models were used to calculate the time-dependent effects of surface shear stresses on thrombi developed *in vitro*. However, as mentioned previously, *in vitro* experiments may not depict realistic physiological conditions observed *in vivo*. Hence, in ref.[45] intravital imaging was used as the basis for calculating shear stresses in the near-thrombus region *in vivo*. However, these simulations were 2D, which may not provide quantitatively accurate descriptions of the 3D thrombus behavior. Moreover, no yield point or time-dependent stress data were reported.

The reason why the latter study was done in 2D is because the thrombus structure changes faster than a 3D confocal microscopy scan can be completed.[30] Moreover, considerable fluorophore bleaching is experienced during 3D image acquisition, even if the scanned thrombus is static. To overcome this problem, here we instead *estimate* the 3D geometry of the clots from the high-speed imaging of their cross-sections. In this manner, our image-driven simulation approach avoids the pitfalls of using either just experiments or just computation alone. To the best of our knowledge, this combination of advanced time lapse imaging and high-fidelity computation provides the first estimate of the thrombi's yield stress *in vivo*.

## 2. Methods

The overall semi-empirical approach used in this work is summarized in **Figure 3.** Namely, the evolution of the clot structures over time is obtained from 2D intravital microscopy of laser-induced injuries in microvasculature of live mice. Then, the 3D shape of the thrombi is estimated by casting assumptions about the relationship between the 2D and the 3D structures (based on our previous observations).[30] Finally, Lattice-Boltzmann Method (LBM) is used to calculate the fluid forces causing the thrombi to deform (i.e. the critical yield stress $\sigma_c$).

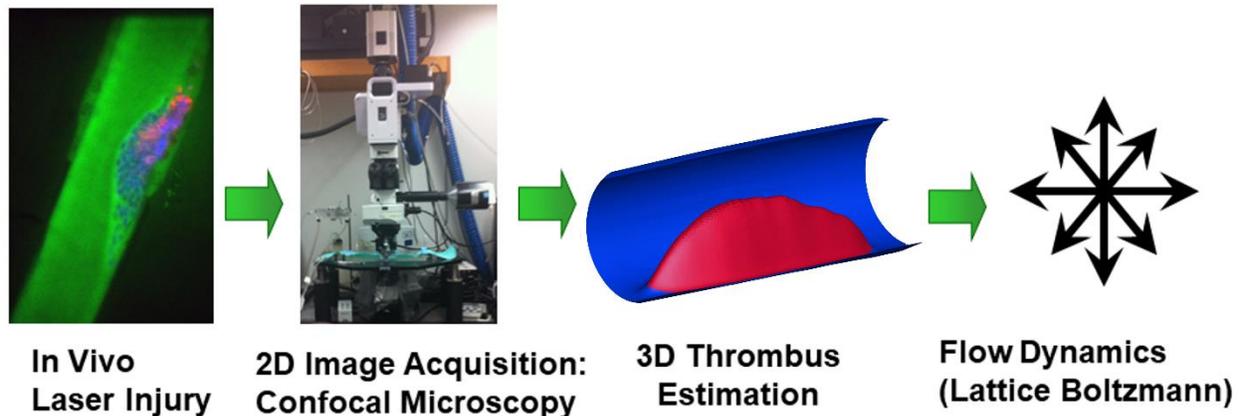

**Figure 3** Process flow diagram for the image-based modeling methodology used in this work. Injury was first induced in the artery of mice to initiate thrombus formation (Blue – indicates activated thrombus core; Red – indicates thrombus shell; Green – represents lumen and interstitial space within the thrombus). Then high-resolution confocal microscopy was used for 2D visualization of the process. Afterwards, a reconstructed 3D thrombus geometry was estimated from the 2D images and imported into the LBM fluid flow solver. Finally, LBM simulation results were used to compute the shear stresses on thrombus surface.

## 2.1. In-Vivo Laser Injury Model and Intravital Microscopy

All of the experiments in this manuscript were performed by the Skip Brass laboratory at the University of Pennsylvania (see Acknowledgements), as described in refs.[20, 46] Briefly, thrombus formation was visualized in the microcirculation of live mice, following procedures previously developed in Ref.[47] In order to image the thrombus structure, intravital microscopy experiments were performed in male C57Bl/6J mice 8-12 weeks of age (Jackson Laboratories, Bar Harbor, ME). Thrombus formation in the mice was induced via laser injury of their cremaster muscle, according to procedures originally developed by Falati et al.[47] Alexa-Fluor® monoclonal antibody labeling kits from Invitrogen (Carlsbad, CA) were used to label antibodies. Anti-CD41 F(ab)$_2$ fragments (clone MWReg30, BD Biosciences, San Diego, CA) were used to visualize platelet surfaces, anti-P-selectin (clone RB40.34, BD Bioscience) was used to visualize degranulated platelets, and caged fluorescein conjugated to albumin[46] was used for lumen and laser injury site visualization. The 2D structure of the thrombus formed due to the laser injury was imaged using confocal microscopy.

The center line maximum velocity through the mouse blood vessels was measured using optical Doppler velocimetry and divided by a factor of 1.6. The latter was done to account for a known artifact of the measuring technique: velocity profiles to appear slightly blunted and non-parabolic due to out-of-focus cells modulating the light intensity signals.[48-50] This procedure yielded an average blood vessel velocity of 4.78 mm/s. The velocimetry measurement was taken in a region away from the thrombus, to ensure that the velocity field was unaffected by the thrombi' presence. More details of the experimental procedures can be found in prior publications [4, 46]. All procedures and protocols were approved by the Institutional Animal Care and Use Committee (IACUC) of the University of Pennsylvania.

## 2.2. 2D Image Acquisition, Post-Processing and 3D thrombus Shape Estimation

Since 3D imaging is too slow to capture clot shape changes, 2D images from the confocal microscope were acquired using SlideBook 5 software (Intelligent Imaging Innovations) with a time interval of 0.619 seconds and a spatial resolution of 0.22 µm/pixel. A total of 300 time images were collected for every experiment. The obtained images were deblurred using a technique reported in Ref.[20] Next, Fiji[51] plug-ins were used to compensate for vibrations due to muscle contractions in the mouse, and air currents surrounding the sample. Specifically, to achieve video stabilization, all intensity channels were superimposed together, and stabilized collectively using either StackReg[52] or Image Stabilizer ImageJ plug-ins;[53] where the choice of plug-in is based on which gave better results for each particular capture. Finally, a custom Matlab code (MathWorks Inc., Natick, MA) was used to further enhance the signal-to-noise ratio of the 2D images and subsequently segment 2D thrombus shapes using standard image processing techniques.

Once the 2D images were post-processed, an in-house Matlab code (MathWorks Inc., Natick, MA) was used to generate *estimated* 3D clot geometries based on the 2D confocal images, at each time point. Specifically, the code assumed that each time-point image represented longitudinal (i.e., *parallel* to the blood flow axis) 2D cross-sections through the center of the *actual* 3D thrombi, at that same instant. This is a good assumption, since the microscope's operator particularly chose an imaging plane such that the thrombi' cross-sections were maximized. A representative grayscale thrombus cross-section from microscopy is shown in **Figure 4-LEFT.** Subsequently, cross-wise (i.e., *perpendicular* to the blood flow axis) parabolic cross-sections were stacked along the "spine" of each clot artificially, to estimate the thrombi's 3D shapes (see **Figure 4-MIDDLE**). The resulting parabolic shape, chosen to represent the outer thrombus perimeter, was selected based on experimentally-observed 3D reconstructions of a static clot.[20] Specifically, the height of each parabola was dictated by the height of thrombus' "spine" at every position along the *longitudinal* direction; while the widths of the parabolas were assumed to be in a constant 2:3 ratio with the parabola heights throughout the thrombus. This ratio was again consistent with experimental observations from static 3D *in vivo* imaging shown in Ref.[20].

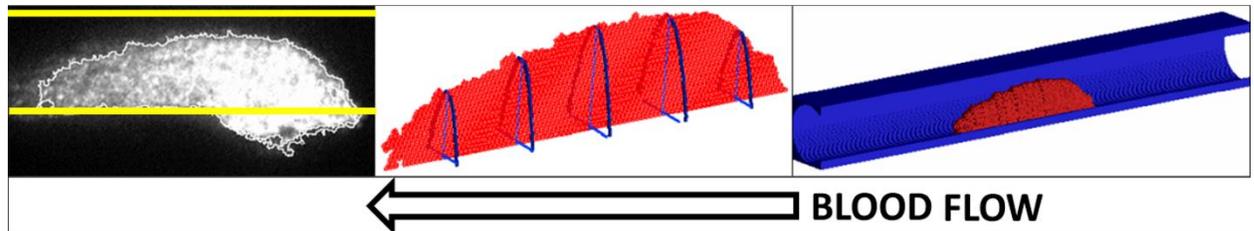

**Figure 4** 3D Thrombus Shape Estimation Proceedure: LEFT – a 2D grayscale image of a thrombus crossection obtained via confocal fluorescent microscopy (yellow lines mark the blood vessel edges); MIDDLE - illustration of how parabola are fit on the edges to generate 3D geometry (red is the 2D cross-section, blue are the parabolas fitted to its "spine"); RIGHT - resulting 3D reconstruction (blue is a pipe fitted around thrombus depicting the blood vessel; red is the estimated thrombus). Pipe diameter and lengths are 0.03685mm and 0.2275mm for this experiment, respectively. White arrow indicates the direction of blood flow.

Finally, the blood vessel walls were manually segmented from the 2D images to measure the lumen diameter. Subsequently, a 3D pipe of the same size was created around the estimated 3D blood clot (see **Figure 4-RIGHT**). The length of the pipe was chosen to be three times the maximum clot length, to avoid entrance effects in fluid flow simulations.

## 2.3. Fluid Flow Modeling: Lattice-Boltzmann Method (LBM)

Convection within the blood vessels and around the 3D thrombi was modeled via LBM. LBM is a numerical technique for simulating mesoscopic fluid flows, which consists of solving the discrete Boltzmann equation.[54-56] In addition to computational advantages e.g., LBM is inherently parallelizable, [57, 58] LBM techniques have been used in a wide spectrum of applications [turbulence, [59] non-Newtonian flows [60-62], and multiphase flows [63]]. Moreover, it has recently gained popularity for modeling biological flows.[64-71]

It was chosen for this work because of the ease with which it handles complex boundary conditions, and due to its efficient parallelizability on supercomputers.[72] Both of these aspects make the method especially attractive for dealing with high-resolution images, such as the ones used in our study. A previously developed custom-written, parallelized, in-house code was used in this work.[20, 64, 66-69, 73-75] The D3Q15 lattice[76] in conjunction with the single-relaxation time Bhatnagar, Gross and Krook[77] collision term approximation was used to perform the simulations in this study. The no-slip boundary condition was applied at wall faces using the "bounce-back" technique.[55] To take advantage of inherent LBM parallelizability, domains were decomposed using message passing interface.[64, 69] The code has been validated for several flow cases for which analytical solutions are available: forced flow in a slit, flow in a pipe and flow through an infinite array of spheres.[64, 68]

The LBM simulations were performed for a total of ten different laser-injury experiments (see

). For each one, a *pseudo-steady state* approach was taken, to update the clot shape based on the successive microscopy images. Namely, flows through geometries representing each individual time step were solved *separately* from each other, as *steady state* simulations. Ultimately, the individual time step results were then concatenated to form a continuous time-series. The pseudo-steady state approach is a good assumption, given than the fluid velocity field changes much faster than do the clot geometries.

**Table 1** Summary of data for the 10 modeled thrombi.

| | Maximum stress (dynes/cm$^2$) | | Maximum | | | | | |
|---|---|---|---|---|---|---|---|---|
| **Blood Vessel Diameter [μm]** | **Constant Pressure Drop** | **Constant Flow Rate** | **CD41 Area[μm$^2$]** | **P-Selectin Area[μm$^2$]** | **CD41 Height [μm]** | **CD41 Volume[μm$^3$]** | **Injury length[μm]** | **Specific Surface Area [μm$^2$/μm$^3$]** |
| 28.30 | 57.12 | 92.68 | 749.89 | 82.98 | 25.90 | 4447.79 | 27.23 | 55838.42 |
| 34.18 | 55.57 | 81.23 | 1580.69 | 137.87 | 36.05 | 10922.70 | 25.63 | 39835.00 |
| 30.44 | 54.27 | 56.16 | 447.91 | 107.58 | 17.62 | 906.50 | 13.88 | 58685.18 |
| 34.71 | 55.83 | 93.32 | 1333.18 | 180.93 | 37.38 | 10881.41 | 36.31 | 50712.37 |
| 36.85 | 51.40 | 82.88 | 1759.63 | 170.45 | 37.91 | 16455.41 | 26.17 | 39182.72 |
| 41.12 | 44.23 | 61.67 | 1313.43 | 183.64 | 40.05 | 9818.34 | 26.97 | 53490.24 |
| 42.19 | 43.43 | 76.78 | 1755.28 | 174.02 | 43.25 | 17243.75 | 30.97 | 54409.11 |
| 51.26 | 33.28 | 33.80 | 2037.72 | 201.61 | 36.85 | 13299.55 | 50.46 | 37391.42 |
| 57.67 | 27.58 | 30.80 | 2297.07 | 173.66 | 43.25 | 17016.10 | 41.39 | 46383.00 |

| 45.92 | 35.43 | 48.89 | 2742.20 | 180.72 | 46.73 | 21660.17 | 59.81 | 43235.64 |

While the fluid dynamic viscosity of blood is known to be shear-dependent,[78, 79] here it is assumed to exhibit Newtonian behavior with a viscosity of µ ≈ 0.03 gr/cm-s.[80, 81] This is considered to be a good assumption due to the low hematocrit typically observed in microvasculature.[82] Given that the reality is likely to be something in-between, two different boundary conditions commonly encountered in literature were modeled: constant pressure drop[25, 44, 83] and constant flow rate.[44, 84] For the latter, the flow is continuously adjusted to ensure that input flow rate is maintained consistently throughout thrombus development. In both cases, the *initial* average lumen blood velocity in LBM was matched to the *in vivo* value of 4.78mm/s, obtained by optical Doppler velocimetry for a comparable diameter blood vessel (as discussed in the experimental methodology Section 2.1). The simulations were considered converged when smallest, average, and highest flow velocities in the whole simulation domain varied by less than 0.01% per LBM 1000 steps. Approximately 60,000 steps were needed for full conversion of each experimental time step. The image-based LBM simulation results were validated against an idealized homogeneous porous media model of a thrombus using a commercial computational fluid dynamics package COMSOL®.

## 2.4. Stress Calculations

Since the thrombi deformation was obtained from imaging, their geometry in the model was assumed to be solid. The fluid-induced shear stresses acting on thrombi surfaces were estimated following a scheme suggested by Porter et al.[65] using the equation (3):

$$\underline{\underline{\sigma}} \approx \mu \left(\frac{1}{2}\right)\left(\nabla U + \nabla U^T\right) \tag{3}$$

Where $\underline{\underline{\sigma}}$ is the shear stress tensor and $U$ is local velocity vector.

Derivatives of the velocity field given in equation (3) were approximated numerically using centered finite difference method (Equations (4) – (6)). The same was done for the partials of $U_y$ and $U_z$. Following this, the symmetric strain matrices were found by adding the 3 × 3 partials matrix for each field location to its own transpose.

$$\frac{\partial U_x(i,j,k)}{\partial x} = \frac{U_x(i+lu,j,k) - U_x(i-lu,j,k)}{2 \times lu} \tag{4}$$

$$\frac{\partial U_x(i,j,k)}{\partial y} = \frac{U_x(i,j+lu,k) - U_x(i,j-lu,k)}{2 \times lu} \tag{5}$$

$$\frac{\partial U_x(i,j,k)}{\partial z} = \frac{U_x(i,j,k+lu) - U_x(i,j,k-lu)}{2 \times lu} \tag{6}$$

Where $lu$ is the length of one side of an element in the LBM model.

The 3×3 partials matrix for each point in the field was added to its transpose to produce a symmetric strain matrix. The eigenvalues of the symmetric matrix are found using the Jacobi method, and largest absolute-value eigenvalue are used to determine shear stress. It is important to note that only stresses due to fluid shear are reported in this study. The upstream wall stress value obtained from stimulation was ~31 dyn/cm$^2$, which is comparable with experimental value of 40 dyn/cm$^2$ from literature.[85] 3D reconstructions generated using Tecplot 360 EX 2017 (Tecplot Inc., Bellevue, WA USA) were used for visualizing velocity and stress distributions acting on and around the thrombi.

## 3. Results

The material properties of thrombi provide a measure of when the clots can potentially become pathological. The goal of this paper is to *estimate* the mechanical strength of the thrombi formed in the microcirculation of live mice. This is done by calculating the stresses induced on the clot by the surrounding blood flow, from LBM simulations based on intravital microscopy images. The hybrid semi-empirical approach helps to overcome limitations of conventional experimental and simulation tools: such as, the inability to do fast 3D fluorescent imaging in case of the former, and the uncertainty in generating the thrombi structures mathematically, in case of the latter.

### 3.1. Nondimensionalization via Data Normalization

Firstly, it was found that although each clot has unique properties (size, formation kinetics, etc.), the thrombo-genesis dynamics look similar for different clots when compared on a non-dimensional scale. Therefore, to normalize all clots to the same scale, we nondimensionalized the experiment time by dividing it by a "characteristic" one, as shown in Equation 7

$$t^* = t/t_{char} \qquad (7)$$

where t$^*$ is the dimensionless time, 't' is the experiment time, t$_{char}$ is some characteristic time.

To define the latter, we hypothesized that time of the "yielding" could be used for this purpose, given that it appears to be a transition that is characteristic to all blood clots. We have also noticed that the yielding coincides with the peak in the blood clot size, as measured by the anti-CD41 marker area in the microscopy images. Hence, we chose the time of the peak clot size as the 't$_{char}$', for simplicity (since it doesn't require performing simulations). The normalization results are shown in **Figure 5**.

Here, the thrombus size (quantified via the anti-CD41 platelet marker area in the fluorescence images) and aspect ratio (a measure of morphology, defined as the clot height divided by clot length, also measured via anti-CD41) are plotted versus the dimensionless time, t$^*$. The ordinate values are also normalized by their respective maxima, as follows

$$A^* = A / A_{max} \qquad (8)$$

where, A$^*$ is the dimensionless clot area, A is the clot area as measured from the anti-CD41 marker's fluorescence, and A$_{max}$ is the peak value of A.

$$AR^* = AR / AR_{max} \qquad (9)$$

where, AR$^*$ is the dimensionless aspect ratio of the clot, AR is the aspect ratio of the clot as measured from the anti-CD41 marker's fluorescence, and AR$_{max}$ is the peak value of AR.

Several observations can be made from **Figure 5**. Firstly, $A_{max}$ occurs at $t^* = 1$, as is expected from the normalization procedure. Secondly, $AR_{max}$ occurs at $t^* = 0.5$, which we define as the *beginning* of the clot's deformation region, because at this point the $AR^*$ (i.e., the clots morphology) begins to experience a significant elongation. The yielding is not instantaneous however, as the clot continues to gain mass beyond $t^* = 0.5$. This is evident from the red curve in **Figure 5**. Finally, the clot growth and deformation settle down at approximately $t^* = 1$. Hence, the two mentioned time points appear to be physiologically characteristic of the thrombo-genesis process.

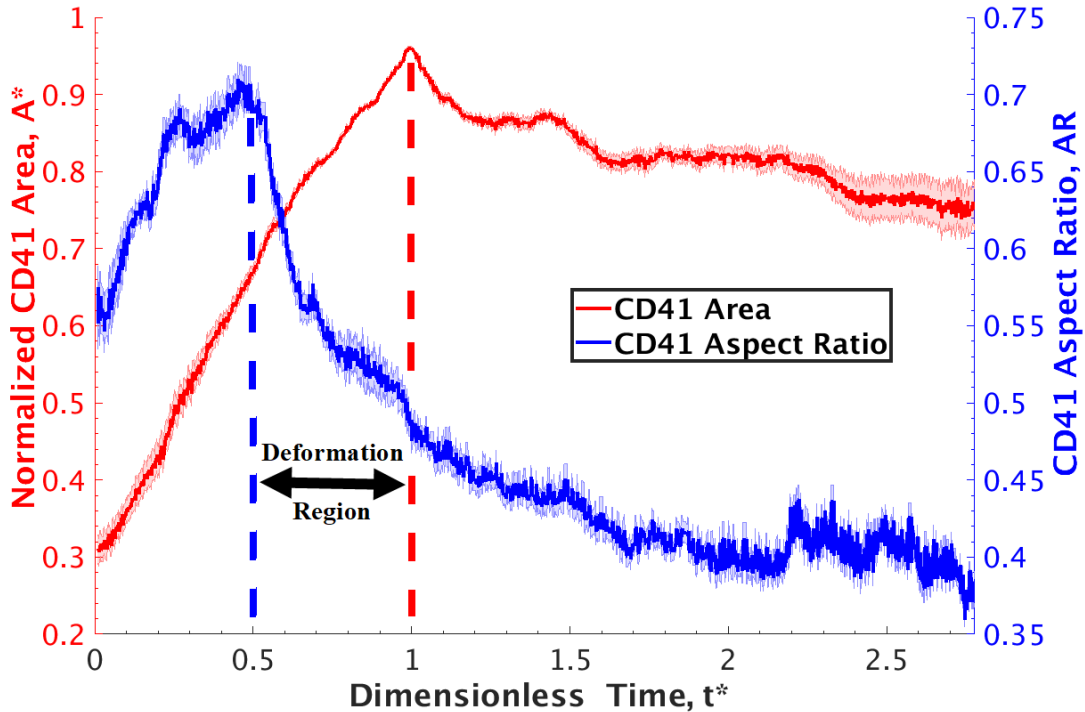

**Figure 5** Data nondimensionalization strategy: Red graph shows the clot's cross-sectional size, quantified based on the anti-CD41 marker area in the microscopy images. Blue graph shows the clot's aspect ratio = height / length. Abscissa is the dimensionless time; whose unity corresponds to the peak CD41 area in the red color curve. Both curves are moving averages with a window of 0.5 and the error bars represent the moving variance for the 10 experiments in
.

### 3.2. Clots Experience Heterogeneous Deformation During Thrombogenesis

Interestingly, the thrombus' inner region (i.e., the "core") does not show a similar trend of deformation. **Figure 6** shows an analysis of the cores' morphology analogous to **Figure 5**. Namely, this figure plots the core's P-selectin area and aspect ratio on the same dimensionless time scale, $t^*$.

Similar to the overall thrombus growth shown in **Figure 5**, the core increases in size up until $t^* = 1$, after which it stays about the same size. Physiologically, the core's growth corresponds to the activation of the platelet mass nearest the injury, similarly to what was shown in **Figure 1**. However, unlike the overall thrombus shape, the core's aspect ratio remains unchanged throughout the whole formation of the clot. From this, it can be concluded that most the clots' deformation occurs in the only thrombus' outer region (i.e., the "shell"), where loosely bound platelets are re-arranged in response to fluid flow stresses.

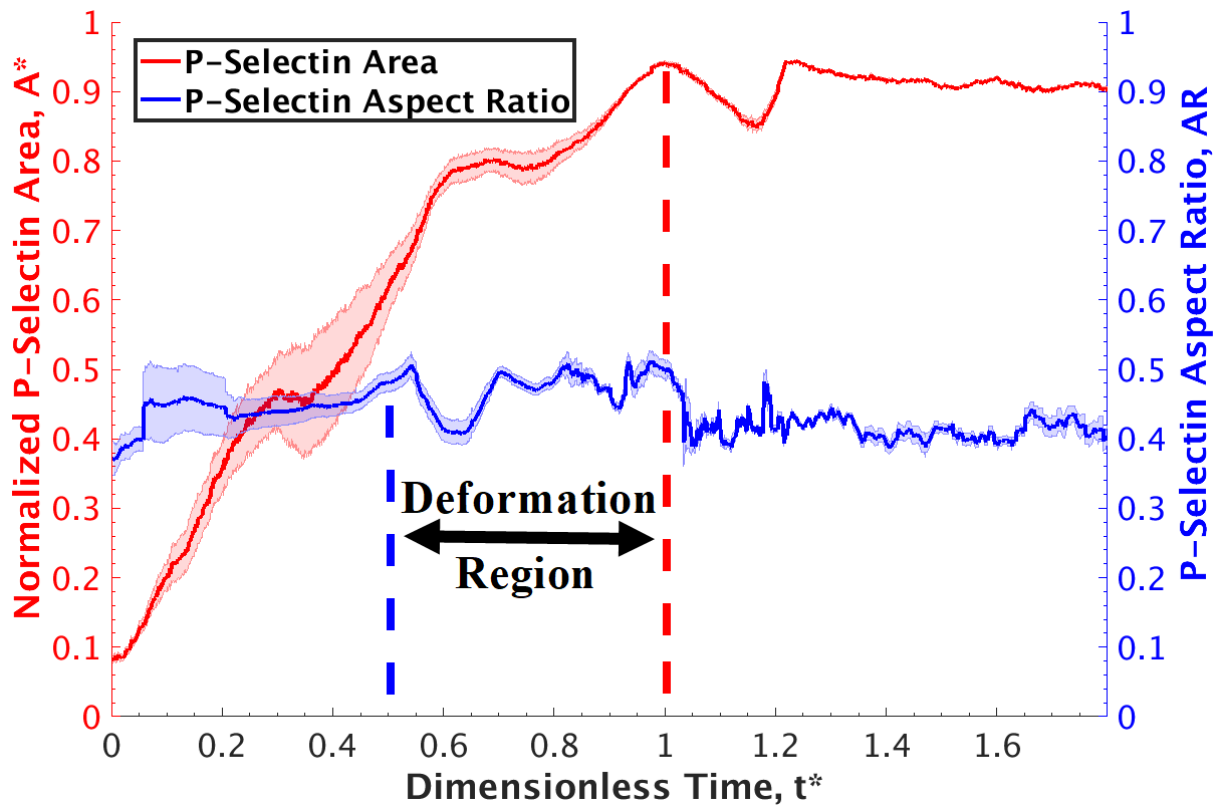

**Figure 6** Changes in nondimensionalized thrombus' core size and morphology, plotted as function of normalized time. Red graph shows the core's cross-sectional size, quantified based on the anti-P-selectin marker area in the microscopy images. Blue graph shows the core's aspect ratio = height / length. Abscissa is the dimensionless time; whose unity corresponds to the peak CD41 area in the red color curve in **Figure 5**. Both curves are moving averages with a window of 0.5 and the error bars represent the moving variance for the 10 experiments in
.

### 3.3. Image-based Modeling of the Fluid-induced Stresses Imposed on the Clots

Since the thrombus partially obstructs blood flow within the vasculature, the thrombus structure experiences forces exerted onto it by the passing fluid. These forces can drive thrombus break-up and embolism. Additionally, platelet adhesion and aggregation [86] and activation [87, 88] are influenced by local shear rate. Hence, we wanted to correlate how the fluid-induced stresses that the thrombi experience due to the flow in the blood vessels correlate to their deformation. To do this, we reconstructed the 3D thrombus shapes, estimated from the intravital microscopy images, in a virtual blood vessel (procedure in **Section 2.2**). Once the clot geometry was obtained, an in-house LBM code was used to calculate the blood velocity field established around the thrombi versus dimensionless time (see **Supplemental Figure 1** and **Video 1**).

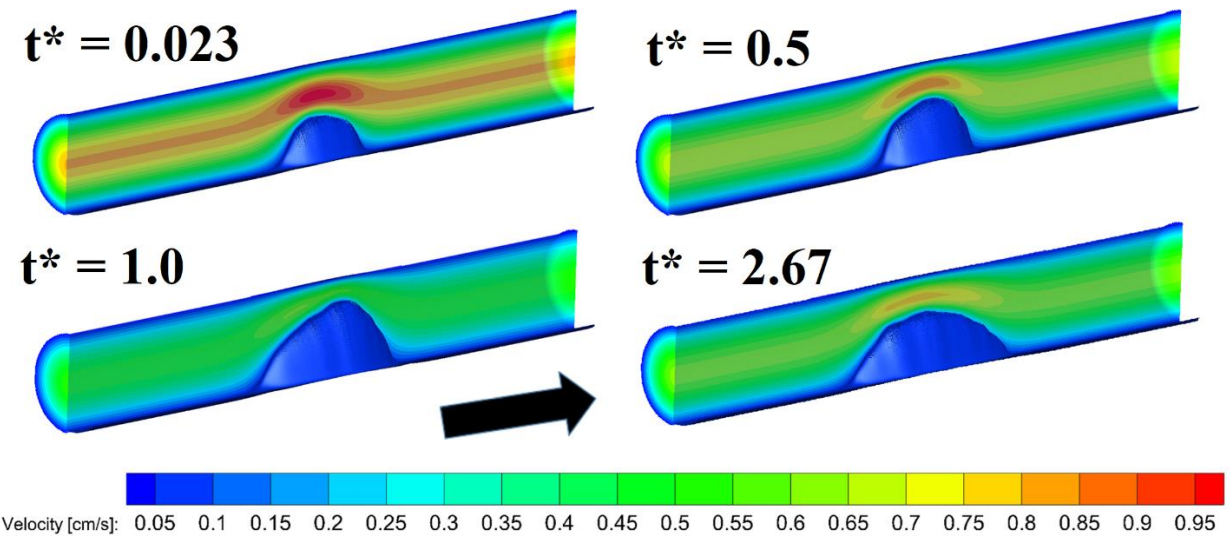

**Supplemental Figure 1** "Heatmap" of the LBM velocity field established in the lumen, calculated using the constant pressure boundary condition. The results are shown at different key time points: $t^* = 0.023$ is an earlier time in the thrombus formation process; $t^* = 0.5$ is the time marking the beginning of deformation/yielding; $t^* = 1$ is the time marking the end of deformation/yielding; $t^* = 2.67$ is a later time, after the thrombus yielding, during which it has assumed its final shape. Black arrow indicates direction of flow.

This figure shows how the velocity changes for a typical thrombus, under the *constant pressure drop* boundary conditions. In this case, the blood flow around the clot *decreases*, as the growing thrombus creates a larger and larger resistance to the flow. This is because at a constant pressure drop, it is more difficult to overcome the growing resistance of the thrombus protruding into the center of the lumen.

Conversely, in the case of the *constant flow rate* boundary condition (not shown), the velocity would *increase* to push the same amount of fluid through a narrower opening in the lumen. As discussed previously in Section 2.3, we performed both types of the simulations, to obtain the upper and lower bounds of the stress experienced by the thrombi. This is because the physiological reality is likely to be something in between the two boundary conditions: namely, a) an injured blood vessel tends to relax in order to avoid occlusion (similar to the *constant pressure drop* case), b) blood may get re-routed through other pathways in the vasculature (similar to the *constant pressure drop* case), and c) the heart may compensate in order to clear the obstruction by pushing the blood harder (similar to the *constant flowrate* case).

Ultimately, the LBM velocity fields for both boundary condition types were used to calculate the fluid-induced stresses experienced in the lumen, using the procedure described in Section 2.4. A representative result for the *constant pressure drop* boundary condition is shown in **Figure 7** and **Video 2**. As expected, the stress distributions change with time, due to the effects that the evolving thrombus structure has on the surrounding blood flow in the lumen. As validation, the thrombus base (nearest the blood vessel wall) experienced stresses comparable to the value of ~31 dynes/cm$^2$, which is expected for an empty blood vessel of a similar diameter (calculated using Hagen–Poiseuille equation).[89] Moreover, these values are also comparable to the experimentally measured values of ~47 dynes/cm$^2$ for similar size arterioles of cat mesentery.[85] Finally, the top part of the thrombus protrudes into the center of the blood vessel, and experiences stresses that are several fold higher than are imposed on its base. Hence, the highest flow forces are acting on the shell of the thrombus, rather than on the core. This is consistent with the shell being the part of the thrombus that experiences the most deformation.

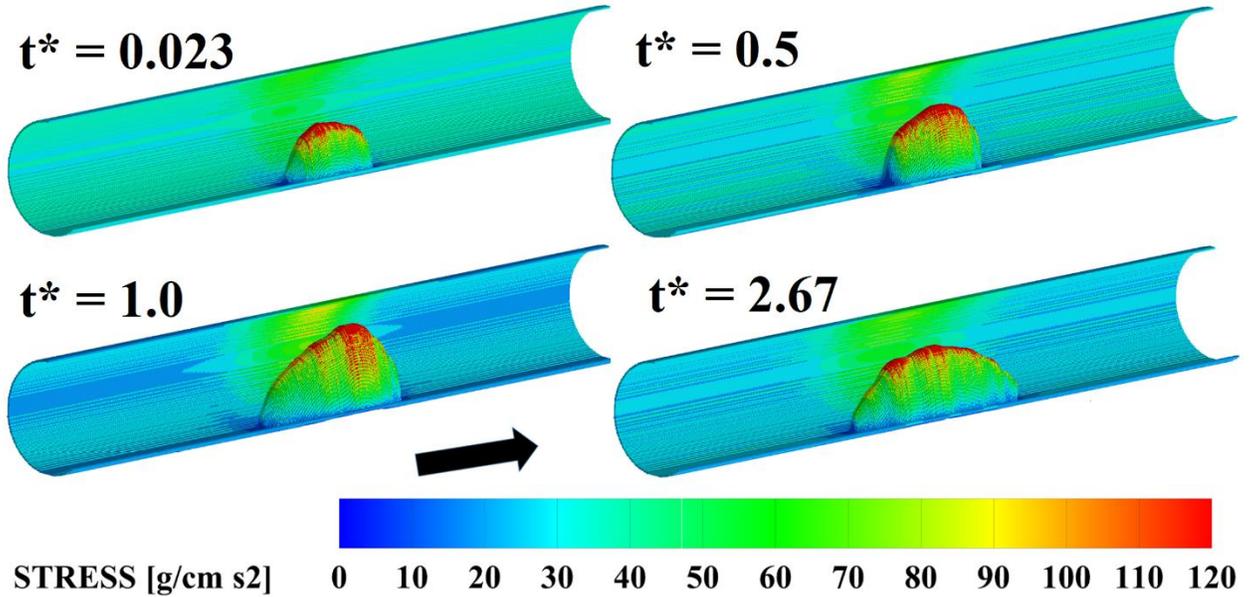

**Figure 7** "Heatmap" of the LBM fluid-induced surface shear stress, caculated using the constant pressure drop boundary condition. The results are shown at different time points. t* = 0.023 is an earlier time in the thrombus formation process; t$^*$ = 0.5 is the time marking the beginning of deformation/yielding; t$^*$ = 1 is the time marking the end of deformation/yielding, when significant mass is transferred to the back of the clot; t$^*$ = 2.67 is a later time after thrombus yielding during which thrombus has assumed its final shape. Black arrow indicates direction of blood flow.

Next, the stresses were plotted as a function of the dimensionless time t*, and normalized as follows:

$$\sigma^* = \sigma / \sigma_{max} \qquad (10)$$

where, $\sigma^*$ is the dimensionless fluid-induced shear stress, $\sigma$ is dimensional fluid-induced shear stress, and $\sigma_{max}$ is the peak value of $\sigma$.

Furthermore, since the simulations are computationally expensive, only every 10$^{th}$ time step out of the total 300 was modeled for each experiment. However, we did solve one of the experiments fully, confirming that the obtained trends would be similar. This confirmation is shown in **Supplemental Figure 2**. From this figure, it is apparent that solving every 10$^{th}$ time step is sufficient to capture the stress trends displayed by the thrombi. Hence, the time-coarsed stress results are shown in **Figure 8**.

In **Figure 8**, the normalized stresses are averaged over the thrombi surfaces, and plotted versus the normalized time, for both the *constant pressure drop* and *constant flow rate* scenarios. In both cases, as the thrombi grow in the blood vessel, the shear stresses imposed on their surfaces initially increase with time. However, the locations of the peaks do not coincide. Instead, they are located approximately at the beginning ($t^* = 0.5$) and the end ($t^* = 1$) of the clot's "deformation region", for the *constant pressure drop* and *constant flow rate* cases, respectively. This again, supports the notion that the two boundary conditions represent extreme cases, while the reality is likely to be something in-between them. Furthermore, it is worthy of noting that in the limit of long time, the stresses decrease roughly to their initial values, while the clot sizes remain nearly constant after peaking. This supports the notion that the thrombi shape rearrangements are driven by drag minimization.

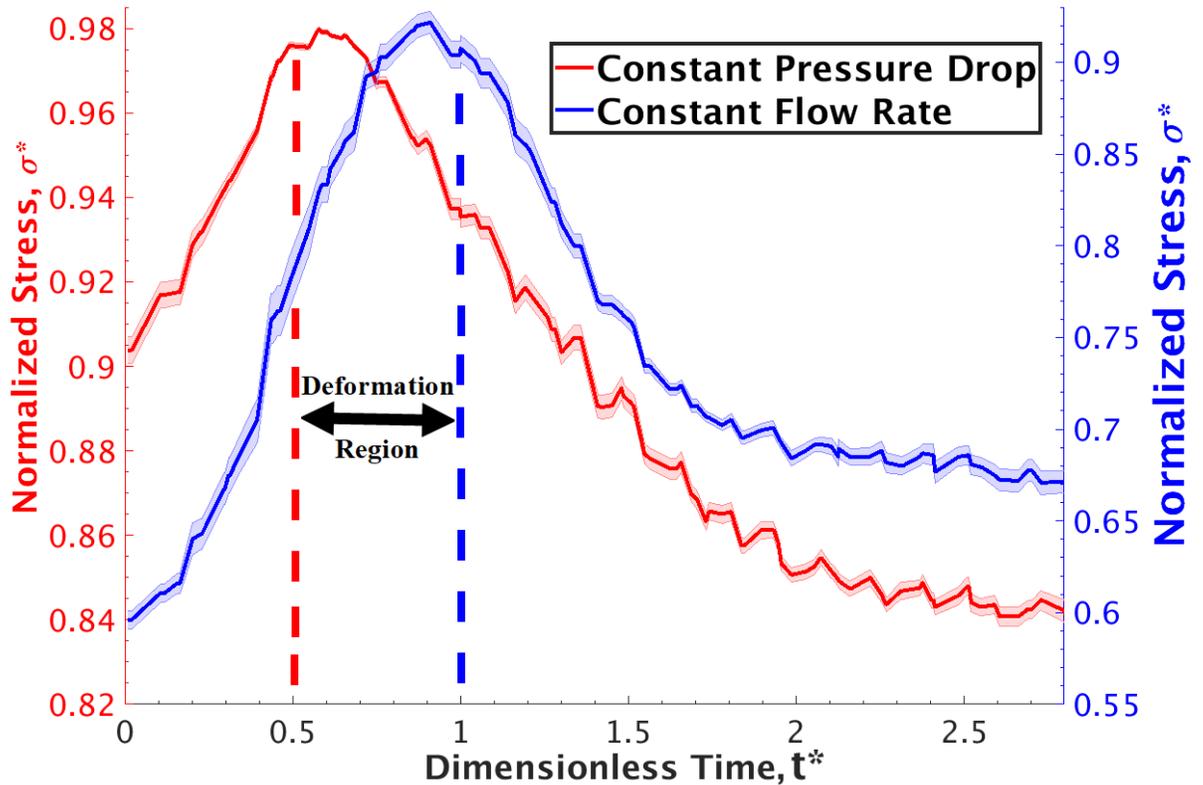

**Figure 8** Normalized stress data, averaged over the thrombus surface, and plotted as a function of normalized time. Blue curve represents *constant flow rate* simulation results. Red curve represents *constant pressure drop* simulation. Abscissa is the dimensionless time; whose unity corresponds to the peak CD41 area in the red color curve in **Figure 5**. Both curves are moving averages with a window of 0.5 and the error bars represent the moving variance for the 10 experiments in

.

Finally, **Table 1** catalogues the $\sigma_{max}$ values for the 10 thrombi modeled in this study. These can be used to recover the actual (i.e., dimensional) stresses from the dimensionless curves in

**Figure 8**. Additionally, the $\sigma_{max}$ results can be extrapolated to other blood vessel sizes. For example, we have observed a strong dependence of the yield stresses on the blood vessel diameter (see **Figure 9**). Although, this may seem surprising at first, it appears that geometric attributes of the blood vessel dictate the peak size of the thrombi. Specifically, the maximum clot volume was found to vary with the blood vessel diameter and injury length (see **Figure 10**). Consequently, it makes sense that the clots found in different blood vessels can withstand varying amounts of stresses, depending on how big or small their structures are.

## 4. Discussion

The yield stress is a critical property representing mechanical strength of the thrombi' material; and as such, it provides a measure of when a blood clot can potentially become pathological. Therefore, quantitative characterization of this parameter is important to public health. However, it is difficult to measure *in vivo* by using either purely experimental or purely computational tools alone. Hence, we applied a semi-empirical framework that combined intravital imaging and flow dynamics simulations, to overcome the technological challenges of the conventional approaches. To the best of our knowledge, this is the first measurement of the thrombi' critical yield stress made *in vivo*. Expectedly, our stress results compare well with both the 2D *in vivo* stress calculations (see Figure 4 in Ref. [45]) and, the 40 dyn/cm$^2$ *in vitro* value for "dynamic" elastic modulus of less activated platelet aggregate (see Table 1 in [33]).

Moreover, we have reported that the thrombogenesis mechanism appears to consist of several discrete regimes common to all clots, despite the underlying uniqueness and complexity of their formation. Consequently, we identified the time of the clot yielding to be a characteristic nondimensionalization parameter, which appears to coincide with both mechanics (e.g., the clots stop growing after the deformation at t* = 1) and biology (e.g. P-selectin expression in the core plateaus at t* = 1) of thrombi formation. Hence, we showed that it can be used to collapse data from multiple injury events onto a single master curve. Furthermore, the consistency of the nondimensionalized trends across the multiple blood vessel injuries gives hope that a uniform theory of thrombo-genesis can be developed, which would be able to describe all clotting events using a single analytical expression.

Finally, we showed that the thrombus core does not change shape appreciably. In contrast, the shell experiences significantly higher fluid-induced stresses, which result in its deformation. This raises the likelihood that the shell is an inherently compromised part of the clot (the high shear stresses may facilitate shell fragmentation due to mechanical disturbance); and that as such, it is responsible for most of the embolism. Hence, it can be concluded that the core is structurally stronger, and that it tethers the overall body of the thrombus to the injury. Consequently, the biological differences between these thrombi regions could mean that it is possible to dissolve just the dangerous part of the blood clot (i.e., the embolizing shell), while leaving the useful one (i.e., the core which seals the injury) intact. In this manner, a new generation of drugs could be developed that would selectively target just the shell; thereby avoiding the dangerous bleeding complications of the medications currently available on the market.

Although our study provides important information regarding the thrombus biomechanics, it is necessary to keep in mind that it is only an *estimate*; and that several assumptions and simplifications had to be made to obtain it. Specifically, the *limitations* of this study are as follows: 1) The clots' 3D shapes were extrapolated from their 2D longitudinal cross-sections, by assuming that thrombi crosswise profiles are parabolic in shape and have a constant width-to-height ratio (see **Figure 4-MIDDLE**). Although this was measured experimentally,[20] a late-stage *static* thrombus was used. Hence, this assumption may not work well for the very early stages (i.e., $t^* < 0.5$) of the thrombus formation (when the clot has a less elongated shape). However, this only concerns the 3D LBM simulations, while the 2D experimental results would not be affected; 2) The pseudo-steady state approach to the simulation necessitated the assumption that the velocity field of the blood flow establishes faster than the clots change their shape; 3) The thrombi are assumed to be impermeable to the fluid flow, since the true nature of their porous structure could not be approximated using our minimalistic approach; 4) We did not model blood escape from the injury site, which could cause the simulated flow field to deviate from the real one; 5) The shape of the blood vessel was assumed to be a straight solid pipe, while in reality it may bend and deform, (especially near the injury site); 6) Finally, the blood was assumed to be a Newtonian fluid for simplicity. However, this is commonly done in similar thrombo-genesis modeling studies.[20, 24, 43, 90] Moreover, inaccuracies incurred due to this assumption are believed to be small, since blood behaves close to a Newtonian fluid when the shear stress is larger than 1 Pa (which is our case).[45]

The future directions for this work include performing similar measurements of the blood flow forces experienced by occlusive and embolizing thrombi (both of which are harder to induce and capture experimentally). Furthermore, we plan to continue working towards developing a universal model of thrombo-genesis. This will be done by correlating other physiological phenomena, such as the dynamics of the blood's escape to the extravascular space, to the dimensionless time scale, $t^*$. Additionally, the "plastic viscosity" (in the second term on the right-hand side of Equation 2) will be measured via image-based modeling). Finally, superior imaging methods will be used to capture the real-time 3D thrombi shape changes, instead of relying on extrapolations based on the 2D cross-sections from conventional confocal microscopy.

## 5. Conclusions

In this study, we have performed an in-depth analysis of intravital microscopy images, showing thrombi development in response to laser-induced injuries in live mouse microvasculature. Based on these results, we were able to conclude that the thrombus core does not change shape appreciably during thrombo-genesis, but its shell does. This implies that there are inherent differences in the material properties of these two regions of the clots. Furthermore, we performed image-based LBM modeling, which allowed us to calculate the fluid-induced shear stresses imposed on the thrombi's surfaces by the blood flow. From these results, we observed that it is the thrombus shell that experiences the highest fluid-induced shear stresses on its surface. A combination of the two results, namely that the shell is both weaker and experiences more deformation, leads to the conclusion that it is the most prone to embolism. The implications of this finding are that a new class of anti-embolic drugs could be developed, which would target the dissolution of the shell selectively, while preventing the risk of severe bleeding (typically associated with the existing antithrombotic medications) by leaving the core of the clots intact. Finally, we have laid down a foundation for a nondimensionalization approach to interpreting thrombogenesis data, with the hope that a uniform theory could be developed through an extension of this procedure. Overall, the findings herein are expected to be beneficial to understanding the process of thrombo-genesis, which is central to heart attacks and strokes that are plaguing the public health today.

## 6. Codes and Data Sharing

The codes used for this work can be found at https://git.njit.edu/rvoronov/lbm. The raw data is too large (> 1 TB) for web repository storage. Instead, it is hosted on supercomputing tape storage at https://portal.tacc.utexas.edu/user-guides/ranch, and is provided upon request.

## 7. Acknowledgements

We would like to thank the Prof. Skip Brass laboratory (including Prof. Timothy J. Stalker and Dr. John D. Welsh) at the University of Pennsylvania's Perelman School of Medicine for performing all the experiments and sharing their microscopy data. Also, we acknowledge that a part of this work was initiated under the guidance of Prof. Scott L. Diamond at the University of Pennsylvania Department of Chemical and Biomolecular Engineering, and Bioengineering Institute for Medicine & Engineering.

The study was financially supported by the Gustavus and Louise Pfeiffer Research Foundation, and in part by NIH R01-HL103419 "Blood Systems Biology" grant and AHA 11POST6890012 Postdoctoral Fellowship.

We also acknowledge that computational resources were provided by the University of Oklahoma Supercomputing Center for Education & Research (OSCER), Texas Advanced Computing Center (TACC) at the University of Texas at Austin; Allocations: TG-BCS170001 and TG-BIO160074 by the Extreme Science and Engineering Discovery Environment (XSEDE)[91], which is supported by National Science Foundation grant number ACI-1548562; and Blue Waters Super Computing cluster at the University of Illinois for granting us access to their high-performance computing facilities and supporting our undergraduate intern financially.

## 8. Disclosure statement

The authors have no competing financial interests to declare.

# **Tables and Table Captions**

# FIGURE CAPTIONS

**Figure 9**  Yield Stress as a function of Blood Vessel Diameter: solid markers indicate LBM calculations, while dashed lines are linear least squares fits through the data.

**Figure 10**  Dependence of clot size on blood vessel diameter and injury length. Red "stems" represent the 10 experimental measurements from this study, while the colored mesh is a fit through these points (meant to serve as a guide for the eye).

# FIGURES

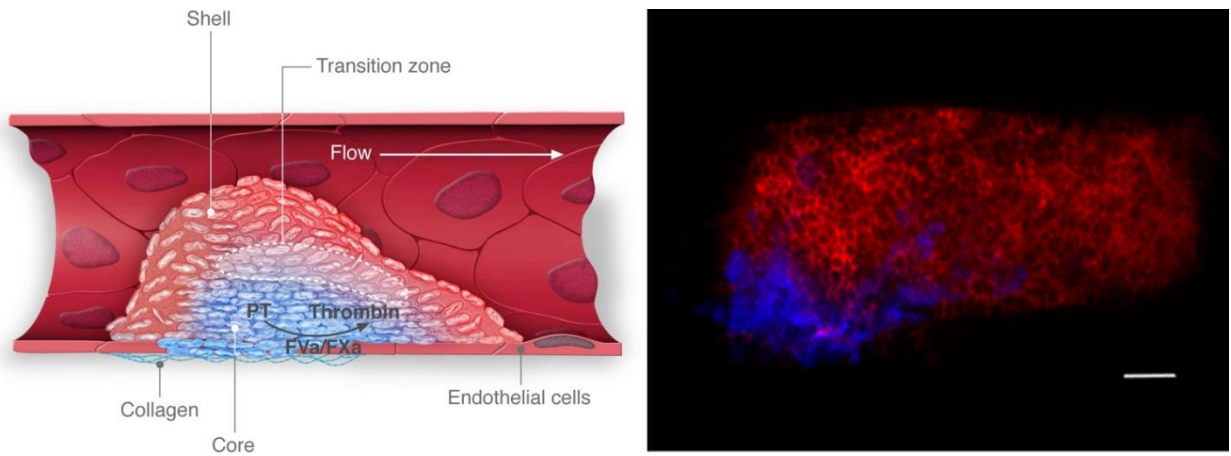

**Figure 1 Kadri et al.**

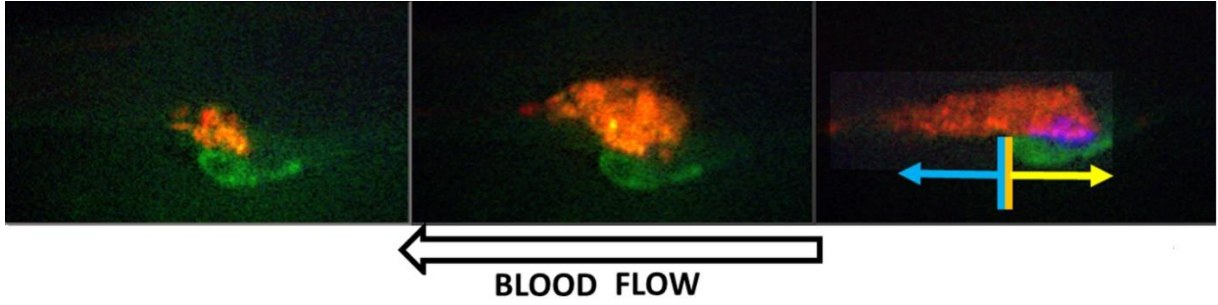

**Figure 2 Kadri et al.**

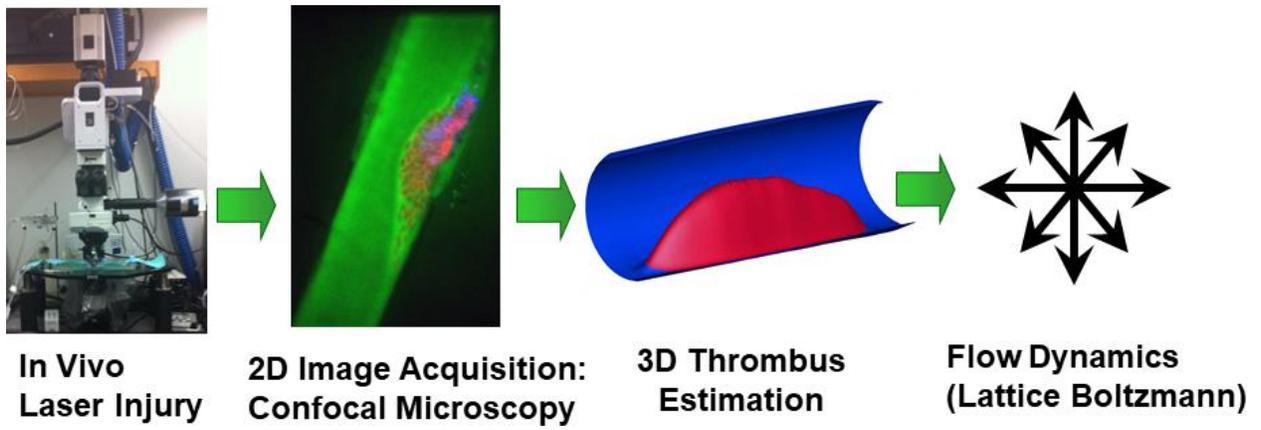

**Figure 3 Kadri et al.**

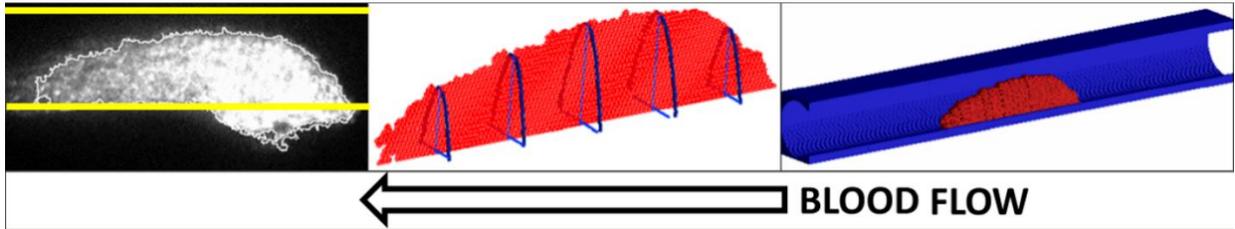

**Figure 4  Kadri et al.**

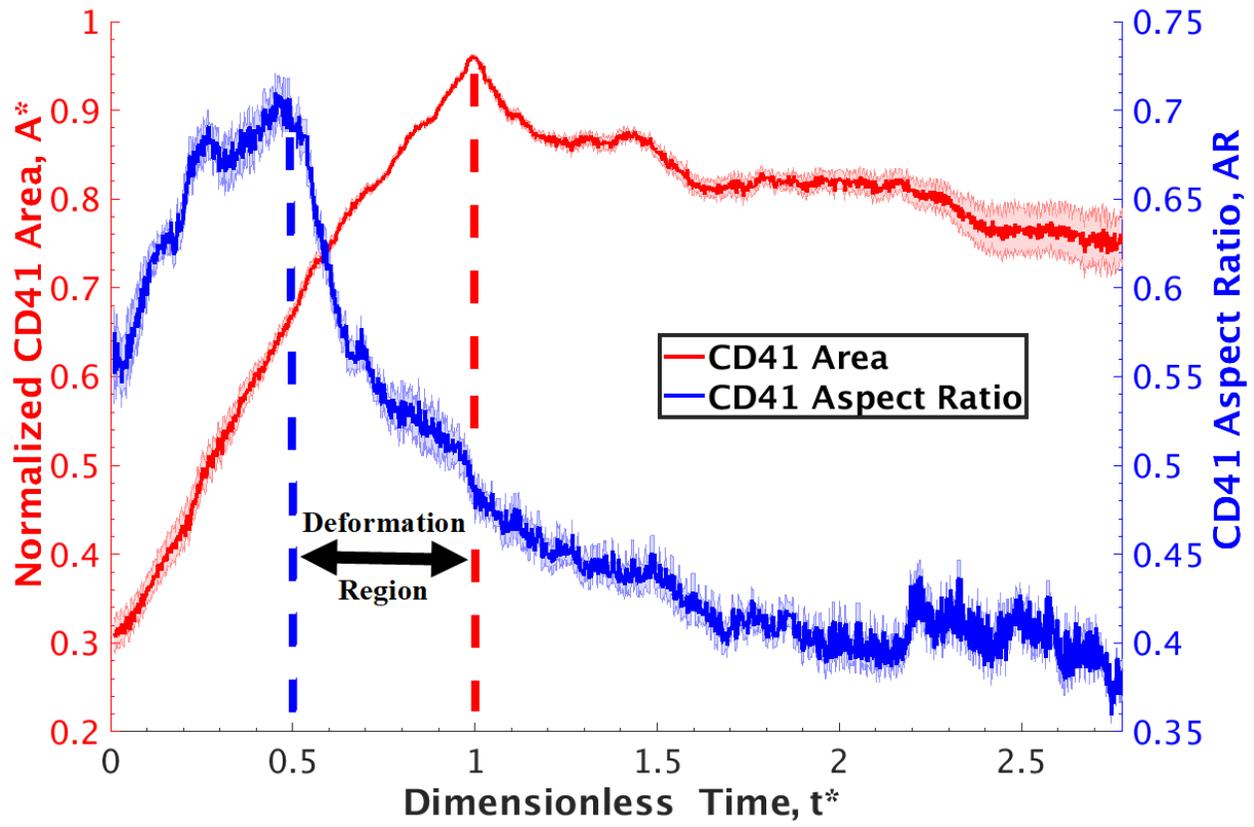

**Figure 5 Kadri et al.**

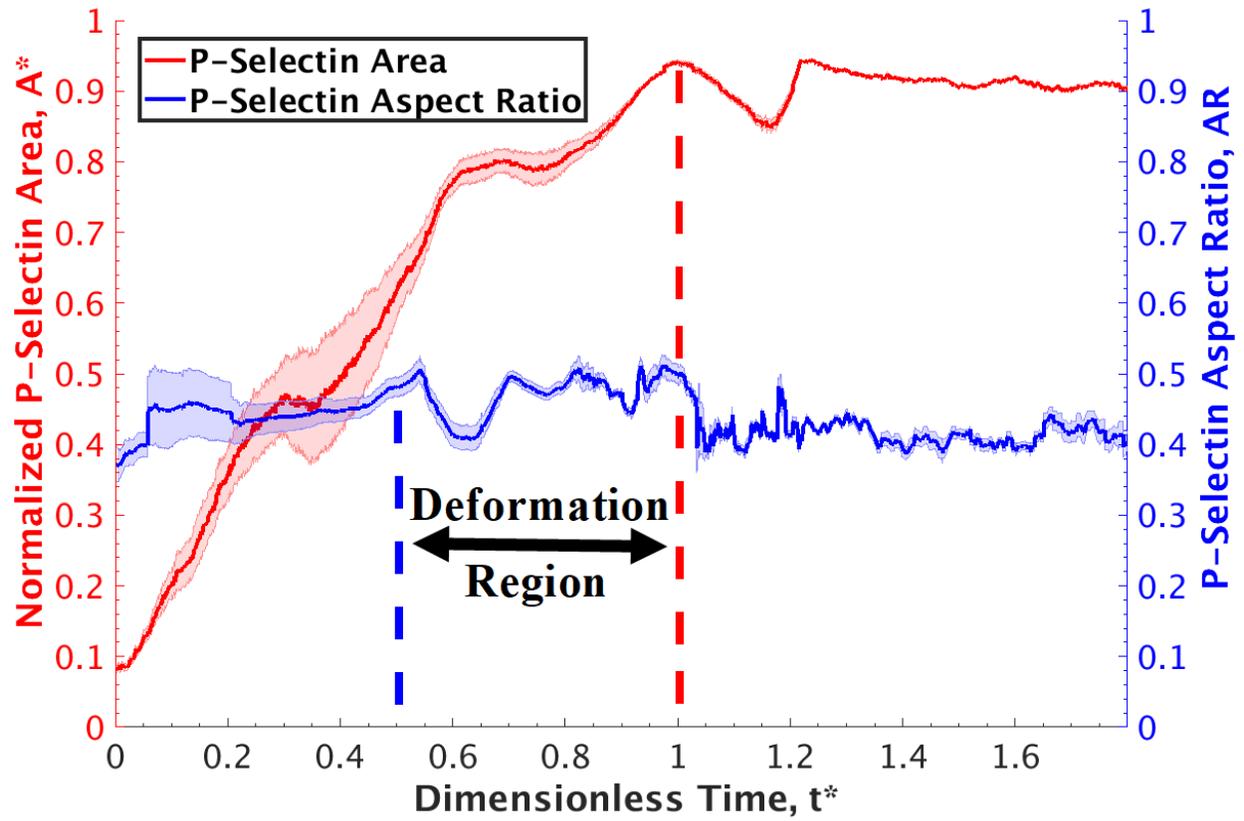

**Figure 6 Kadri et al.**

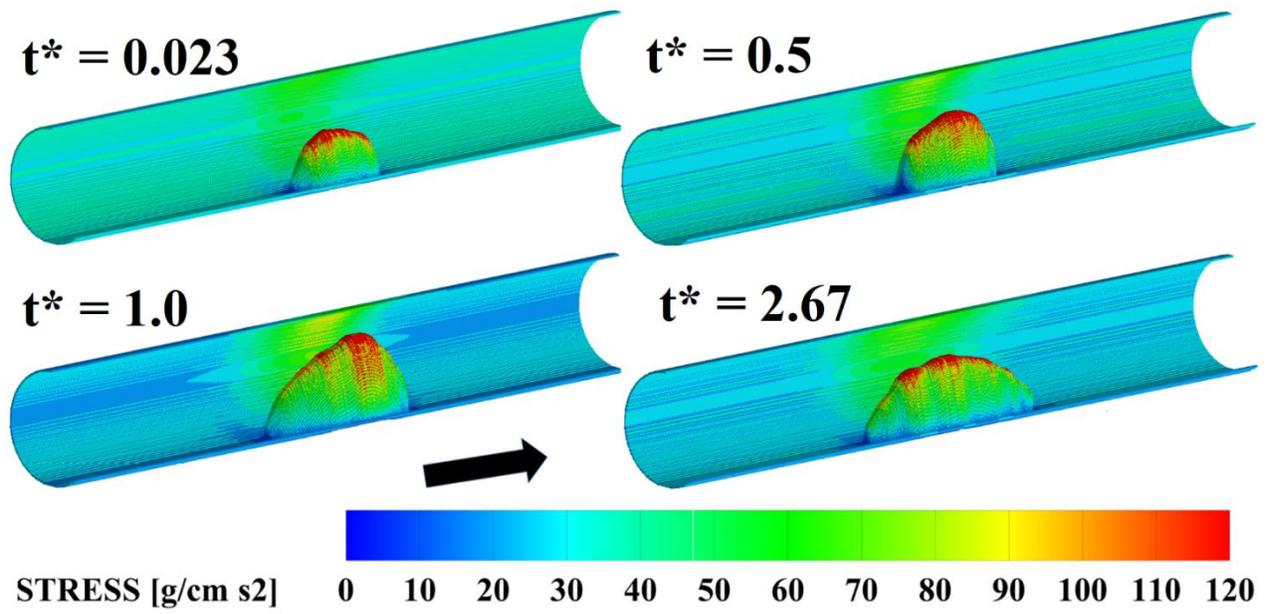

**Figure 7   Kadri et al.**

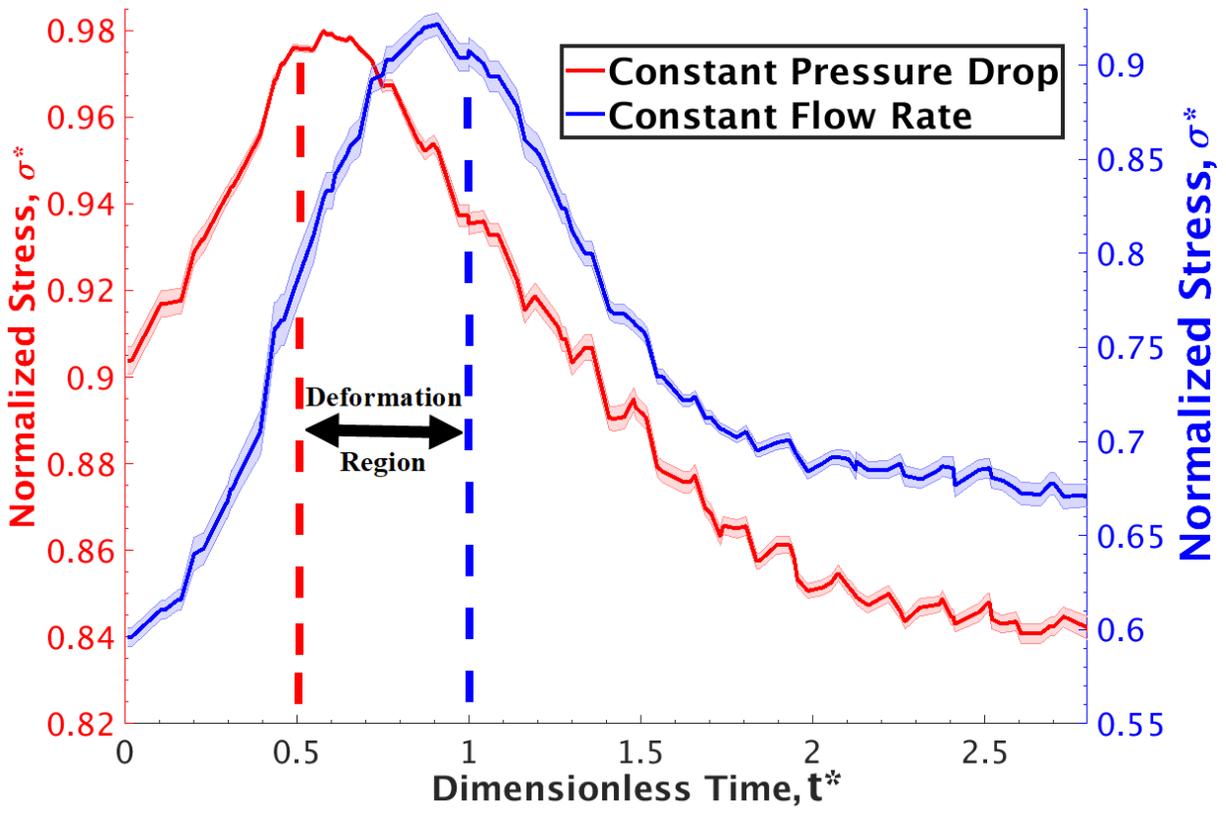

**Figure 8 Kadri et al.**

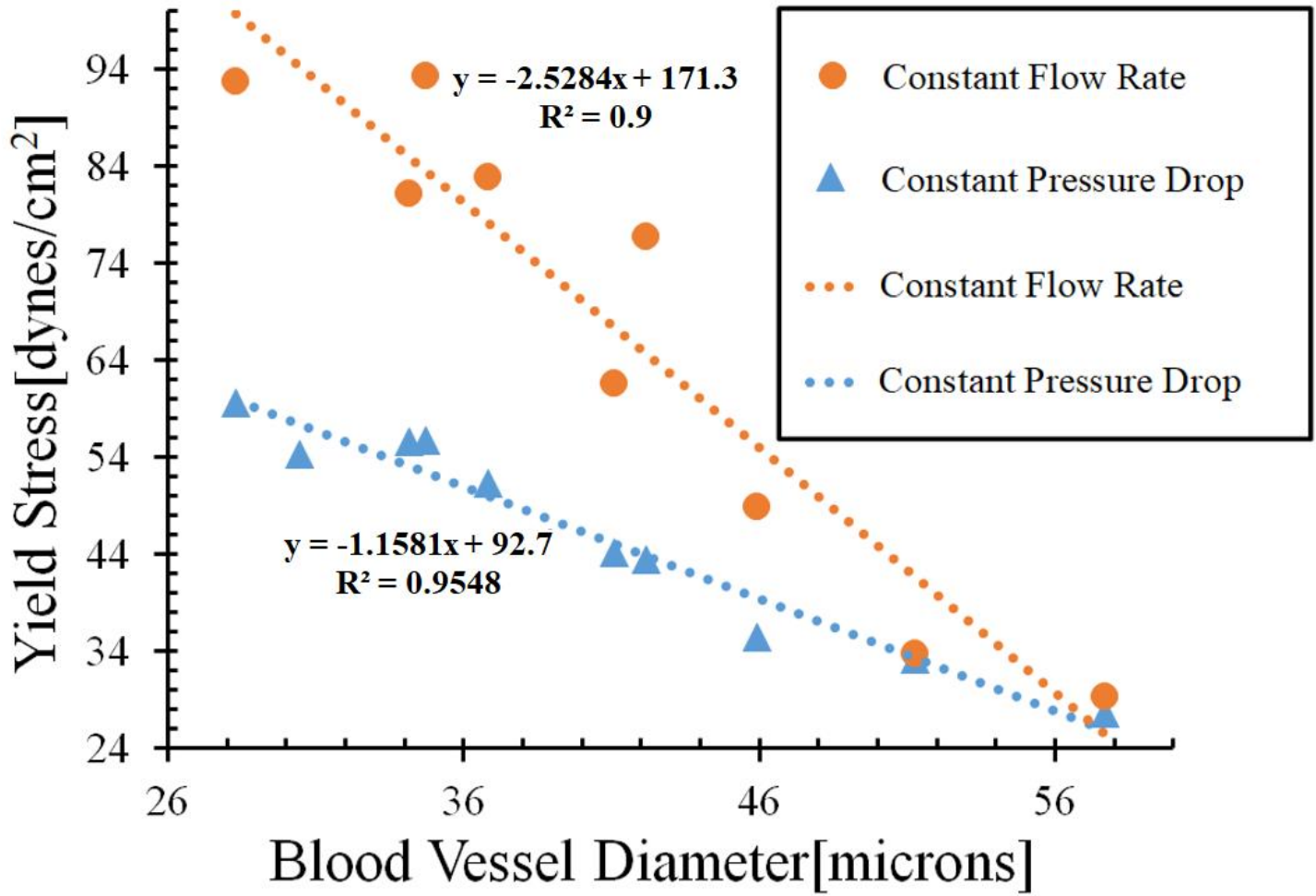

**Figure 9 Kadri et al.**

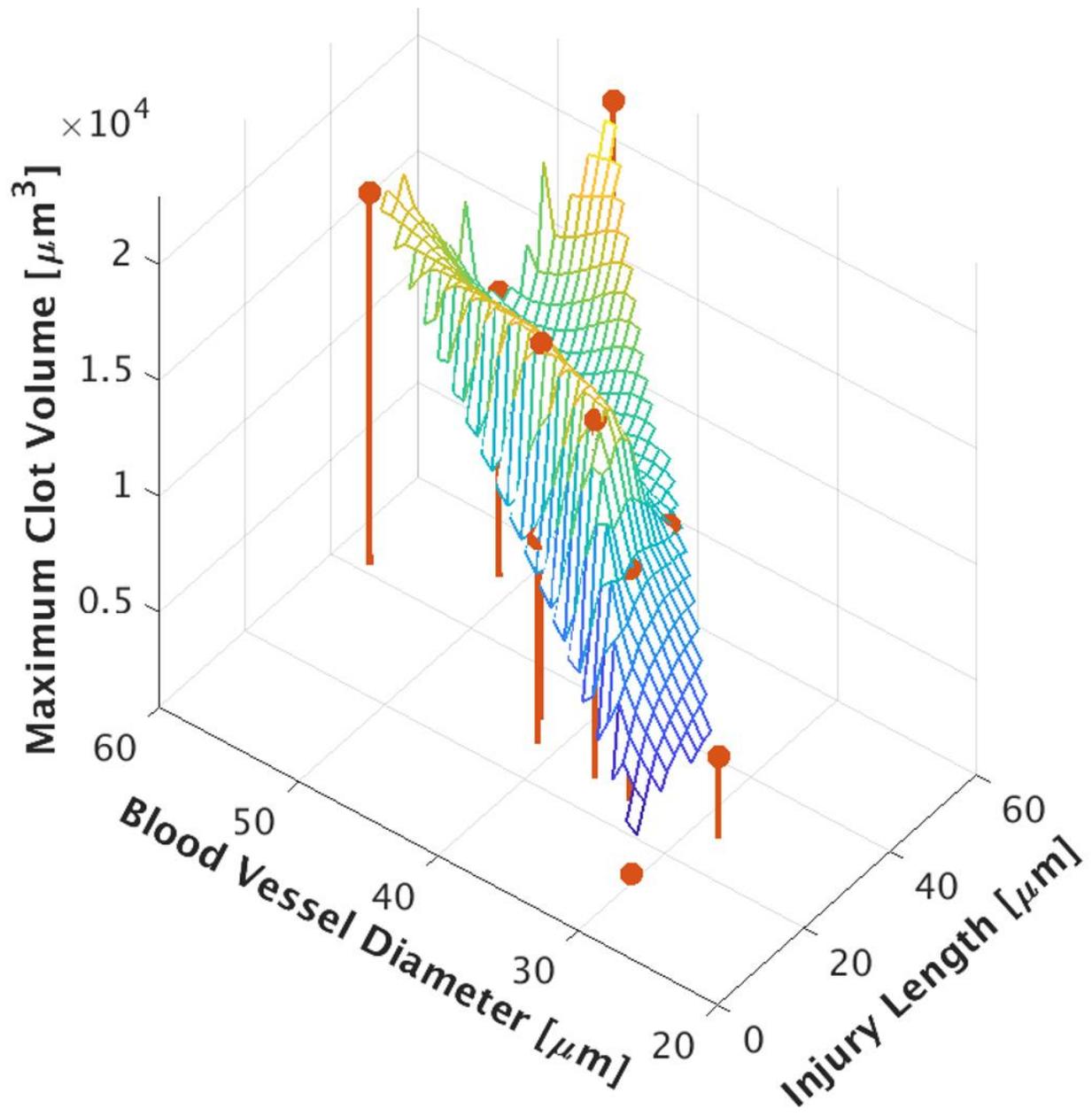

**Figure 10 Kadri et al.**

# VIDEO CAPTIONS

**Video 1**   Animation of 3D LBM results showing how blood clot formation affects the *velocity* field in the lumen of a representative blood vessel. The calculation was carried out using the pressure boundary condition. Blood vessel diameter is 34.18µm.

**Video 2**   Animation of 3D LBM results showing how blood clot formation affects the *fluid-induced stress* field in the lumen of a representative blood vessel. The calculation was carried out using the pressure boundary condition. Blood vessel diameter is 34.18µm.

# SUPPLEMENTAL FIGURE CAPTIONS

**Supplemental Figure 2**  Comparison of LBM stress results for a representative laser-injury induced thrombus, solved for every time step (N=300), versus only every $10^{th}$ one (N = 30). The results agree with each other.

# SUPPLEMENTAL FIGURES

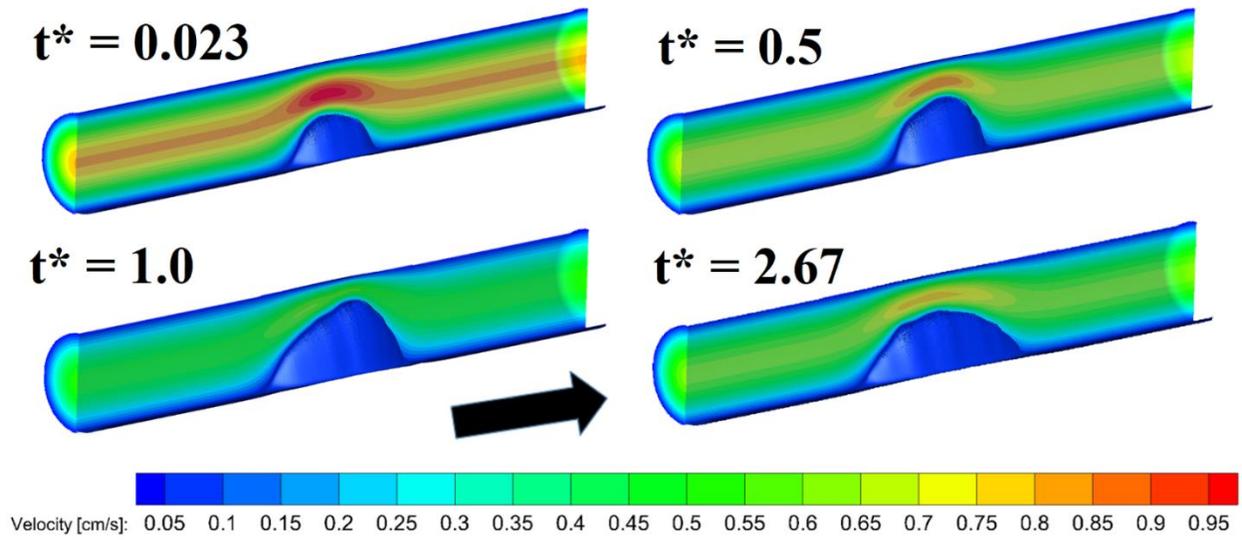

**Supplemental Figure 1 Kadri et al.**

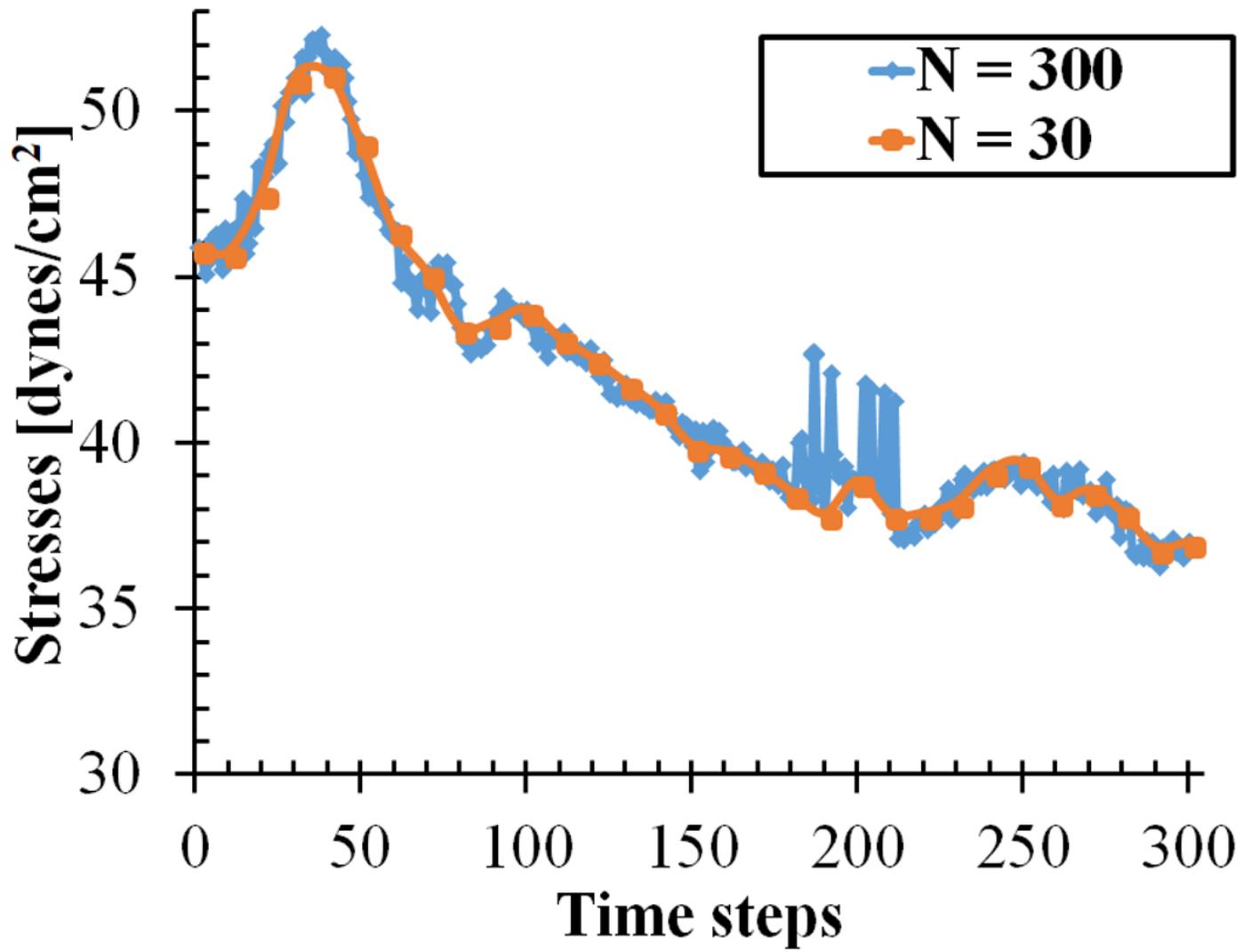

**Supplemental Figure 2 Kadri et al.**